\documentclass[aps,english,prx,floatfix,superscriptaddress,tightenlines,twocolumn]{revtex4-2}
\usepackage[colorlinks=true,linkcolor=blue,citecolor=blue,urlcolor=magenta]{hyperref}
\usepackage{amsfonts,mathrsfs,amsmath,amsthm,amssymb,bbold}                   
\usepackage{epsfig}
\usepackage{bm}
\usepackage{xfrac}
\usepackage{tabulary}
\usepackage{color}
\usepackage[dvipsnames]{xcolor}
\usepackage{graphicx}
\usepackage{ulem}

\newcommand{\la}{\langle}
\newcommand{\ra}{\rangle}

\begin{document}  
\title {\textbf{Squeezing-induced quantum-enhanced multiphase estimation}}

\author{ Le Bin Ho} 
\thanks{Electronic address: binho@fris.tohoku.ac.jp}
\affiliation{Frontier Research Institute 
for Interdisciplinary Sciences, 
Tohoku University, Sendai 980-8578, Japan}
\affiliation{Department of Applied Physics, 
Graduate School of Engineering, 
Tohoku University, 
Sendai 980-8579, Japan}

\date{\today}

\begin{abstract}
We investigate how squeezing techniques can improve the measurement precision in multiphase quantum metrology. While these methods are well-studied and effectively used in single-phase estimations, their usage in multiphase situations  has yet to be examined. We fill this gap by investigating the mechanism of quantum enhancement in the multiphase scenarios. Our analysis provides theoretical and numerical insights into the optimal condition for achieving the quantum Cram{\'e}r-Rao bound, helping us understand the potential and mechanism for quantum-enhanced multiphase estimations with squeezing. This research opens up new possibilities for advancements in quantum metrology and sensing technologies.
\end{abstract}
%
%
\maketitle

{\it Introduction.}---
%
%
%
%
Squeezing in quantum metrology is a technique that manipulates quantum systems to enable more precise measurements beyond classical limits. For instance, squeezed light can enhance the sensitivity of laser interferometers \cite{PhysRevLett.104.103602,SCHNABEL20171}, leading to advancements in gravitational wave detection \cite{PhysRevLett.129.121103, Aasi2013,PhysRevLett.110.181101}, quantum imaging \cite{10.1117/12.2187670}, and quantum lidar \cite{GALLEGOTORROME2024100497,Reichert2022}.
On the other side, collective spin squeezing is crucial for quantum-enhanced precision in Ramsey interferometers \cite{RevModPhys.90.035005}, which are used in atomic clocks \cite{RevModPhys.87.637} and magnetometers \cite{PhysRevLett.109.253605,PhysRevLett.113.103004}. Therein, nonlinear 
transformations
like one-axis-twisting (OAT), two-axis-twisting (TAT), and twist-and-turn (TNT) have been utilized to enhance metrological estimation \cite{PhysRevLett.122.090503}. The OAT has been applied for quantum-enhanced metrology in echo protocols \cite{PhysRevA.94.010102,PhysRevLett.116.053601}, while the TAT has been employed for robust detection-noise interferometry \cite{PhysRevA.97.043813,PhysRevA.98.030303}. These methods, combined with  the interaction-based readout, improve the precision for single-phase estimations \cite{PhysRevLett.119.193601,PhysRevA.97.053618,PhysRevA.98.012129, 
PhysRevA.98.030303}. 
It demonstrates that by twisting and turning, quantum states undergo squeezing and phase shifts, resulting in enhanced sensitivity.
These squeezing transformations have been experimentally realized using Bose-Einstein condensates \cite{PhysRevA.92.023603,Sorensen2001,Riedel2010,PhysRevLett.87.170402,Gross_2012}.

Although the mechanism of squeezing for quantum-enhanced metrology in single-phase estimations is well understood, its application to multiphase estimations remains unexplored. Recent attempts to incorporate squeezing into sensor networks for multiphase estimations \cite{Gessner2020} and using variational squeezing optimization \cite{Le2023,PRXQuantum.4.020333} have been reported. However, these approaches do not fully elucidate the underlying mechanism of quantum enhancement. 

In this work, we explore how nonlinear spin squeezing enhances multiphase estimations. We focus on a scenario where a three-dimensional magnetic field interacts with an ensemble of $N$ identical two-level systems. 
The precision 
is bounded by the quantum Cram{\'e}r-Rao bound (QCRB), which is attainable for single-phase estimations but remains unsaturated for multiphase estimations. 
Previous studies have claimed the saturation of the QCRB for multiphase estimation, however they did not provide the mechanism behind it \cite{PhysRevLett.111.070403,PhysRevLett.116.030801}. Here, we clearly explain 
the mechanism behind this saturation by examining the sysmetry in ensemble systems.
We begin by discussing the saturation of the QCRB for an ideal quantum state characterized by a {\it multi}-GHZ entanglement state, which includes GHZ states in all three spatial directions. We then investigate a realistic case where the quantum state is a single GHZ entanglement in a certain direction. We elucidate the mechanism behind quantum-enhanced precision in this context and assess the impact of noise. Our findings contribute to a better understanding of quantum-enhanced mechanisms in metrology and facilitate the development of quantum sensors and imaging technologies.

\begin{figure*} [t!]
\includegraphics[width=\textwidth]{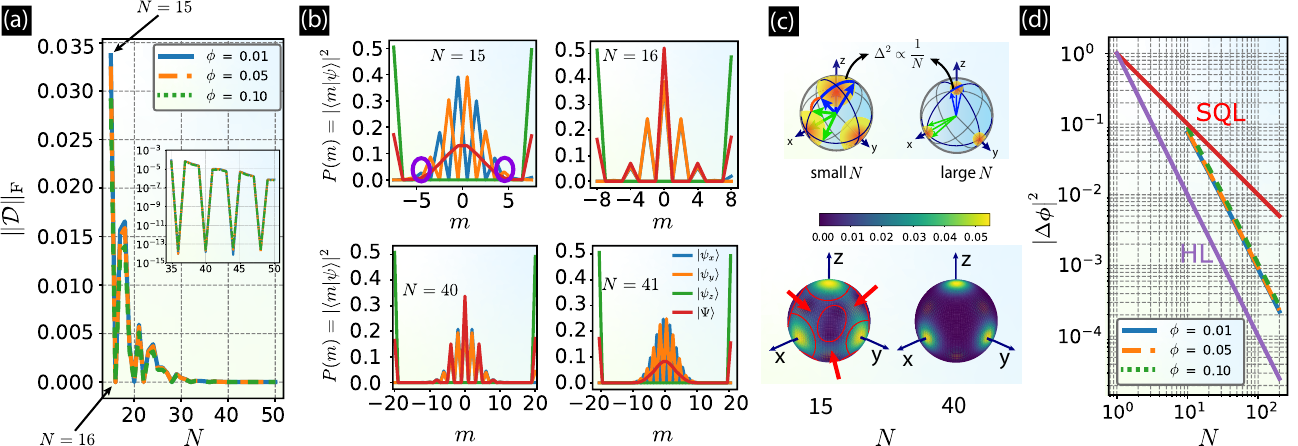}
\caption{
(a) Plot of $\lVert \mathcal{D} \rVert_{\rm F}$ as a function of $N$ for different $\phi$, where $\phi = \phi_x = \phi_y = \phi_z$. $\lVert \mathcal{D} \rVert_{\rm F}$ remains below $10^{-4}$ for $N \geq 35$.
Inset: $\lVert \mathcal{D} \rVert_{\rm F}$ at $N$ from 35 to 50.
(b) The probability $P(m) = |\la m|\psi\ra|^2$, for $|\psi\ra\in
\{|\psi_x\ra, |\psi_y\ra, |\psi_z\ra, |\Psi\ra\}$.
The asymmetry is indicated by purple circuits, and it becomes negligible for large $N$.
(c) (upper) Illustration of spin fluctuation for small and large $N$, where the fluctuation scales inversely with $N$. Here, only one component of each $|\psi_\mu\ra$ is depicted, where the full visualization is shown in the lower of (c).
(c) (lower) Visualization of the Husimi distribution function $Q(\theta, \varphi) = |\la\Psi|\theta, \varphi\ra|^2$ for $N = 15, 16, 40$ and 41. 
(d) Plot of total variance $|\Delta\bm\phi|^2$ as a function of $N$, 
compared with the standard quantum limit (SQL) and Heisenberg limit (HL).
}
\label{fig1}
\end{figure*} 

{\it Quantum-enhanced with a multi-GHZ probe state}.--- 
We examine a 3D vector field 
$\bm \phi = (\phi_x, \phi_y, \phi_z)^\intercal$
that requires estimation.
This field interacts with a probe of $N$ spin-1/2 particles
through a Hamiltonian
\begin{align}\label{eq:H}
H(\bm\phi) = \bm\phi\cdot\bm J =  \sum_\mu \phi_\mu J_\mu,
\end{align}
where $\bm J = (J_x, J_y, J_z)$ is an angular momentum, and 
$J_\mu = \frac{1}{2} \sum_{j=1}^N \sigma_\mu^{(j)}$, 
$\mu = \{x, y, z\}$.
Inspired by single-phase estimations, where the probe state is prepared in a superposition of the maximum and minimum eigenstates of the Hamiltonian $H$ \cite{PhysRevLett.96.010401}, we consider a multi-GHZ probe state $|\Psi\ra$ as
\begin{align}\label{eq:psi}
|\Psi\ra = \dfrac{1}{\mathcal{N}}
\bigl(|{\psi_x}\ra + |\psi_y\ra + |\psi_z\ra\bigr),
\end{align}
where $\mathcal{N}$ is the normalization constant,
$|\psi_\mu\ra = 
\bigl(|\lambda^{\rm max}_\mu\ra
+|\lambda_\mu^{\rm min}\ra\bigr)/
\sqrt{2}$ is a single GHZ component.
Here, $|\lambda^{\rm max}_\mu\ra$ and 
$|\lambda^{\rm min}_\mu\ra$ are two eigenstates
of $J_\mu$ corresponding to the maximum 
and minimum eigenvalues $\lambda^{\rm max}$
and $\lambda^{\rm min}$, respectively. 
This approach has demonstrated 
Heisenberg scaling in the noiseless case
\cite{PhysRevLett.116.030801}.

In this context, the quantum Fisher information matrix (QFIM) gives
(see App.~\ref{appA})
\begin{align}\label{eq:QFIM}
\mathcal{I} = 
4{\rm Re}\bigl[\la\Psi|\bm A^\intercal\bm A|\Psi\ra
-\la\Psi|\bm A^\intercal|\Psi\ra
\la\Psi|\bm A|\Psi\ra\bigr],
\end{align}
where
$
\bm A = \int_0^1du\ e^{iu H(\bm\phi)}\bm J
e^{-iu H(\bm\phi)}
$
are Hermitian operators \cite{10.1063/1.1705306,PhysRevA.90.022117,PhysRevLett.116.030801,Ho2023,PhysRevA.102.022602}.

The performance of an unbiased estimator is determined by the covariance matrix $\mathcal{C}(\bm\phi)$. This matrix has an ultimate lower bound 
known as the quantum Cram{\'e}r-Rao bound (QCRB) \cite{helstrom1967quantum},
i.e., $M\mathcal{C}(\bm\phi) \ge \mathcal{I}^{-1}$, where $M$ denotes the repeated measurements \cite{doi:10.1142/S0219749909004839}.
%
The QCRB 
is attainable
in single-phase estimations \cite{PhysRevLett.96.010401}.
However, to achieve this bound in multiphase estimations, 
it is necessary (but not sufficient) that
$ {\rm Im}[\la\Psi|\bm A^\intercal\bm A|\Psi\ra] = 0$
(see App.~\ref{appC}).
To quantify 
this necessary condition in our case, 
we define a matrix $\mathcal{D}$ as
\begin{align}\label{eq:condition}
\mathcal{D} = 
{\rm Im}[\la\Psi|\bm A^\intercal\bm A|\Psi\ra],
\end{align}
and  derive the Frobenius norm 
$\lVert \mathcal{D} \rVert_{\rm F} 
= \sqrt{\sum_{\mu\nu} |\mathcal{D}_{\mu\nu}|^2}$.
When $\lVert \mathcal{D} \rVert_{\rm F} = 0$, it 
implies that ${\rm Im}[\la\Psi|\bm A^\intercal\bm A|\Psi\ra] = 0$
or in other words, the QCRB in multiphase estimations is attainable.

In Fig.~\ref{fig1}a, 
we show $\lVert \mathcal{D} \rVert_{\rm F}$ as a function of $N$, 
with $\phi_x = \phi_y = \phi_z = \phi$. 
The numerical results indicate that 
$\lVert \mathcal{D} \rVert_{\rm F} \approx 0$ for large $N$, such as,
$\lVert \mathcal{D} \rVert_{\rm F} \propto 10^{-4}$ to $10^{-14}$ 
for $N = 35$ to 50, as shown in the inset Fig.~\ref{fig1}.
This result can be explained by the symmetry of the wavefunction $|\Psi\rangle$ as follows.

\begin{figure*} [t!]
\includegraphics[width=\textwidth]{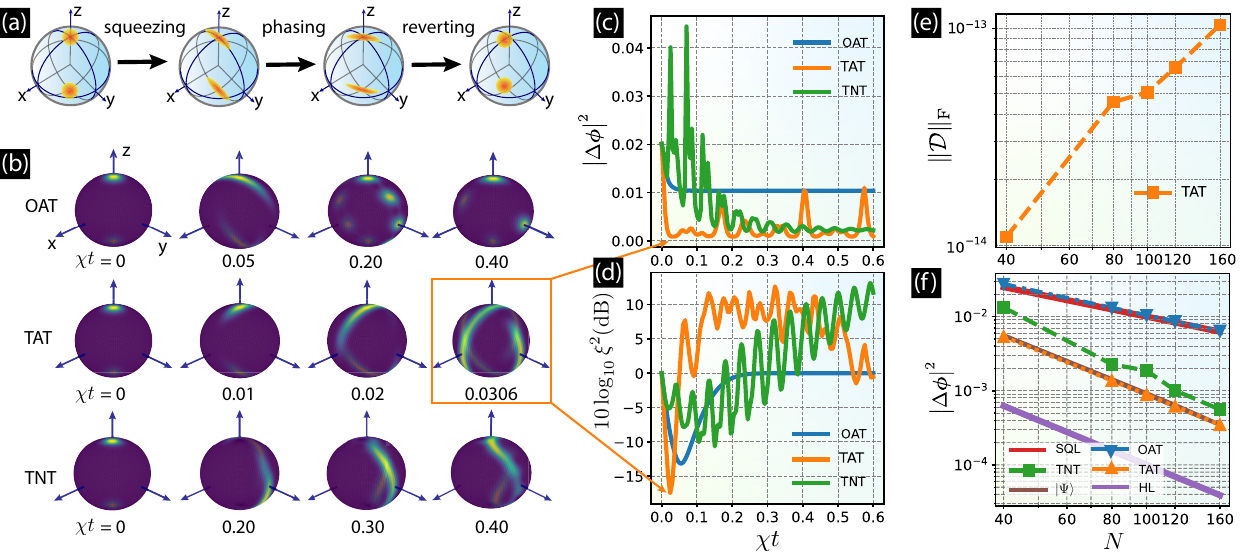}
\caption{
(a) A metrology approach starts by preparing a GHZ state in the $z$ direction, $|\psi_z\ra$, followed by a squeezing transformation, phasing, and reverting transformations. This sequence yields a quantum state containing all necessary information for phase estimation.
(b) Visualization of the Husimi distribution for OAT, TAT, and TNT.
(c) Plot of $|\Delta\bm\phi|^2$ as a function of $\chi t$ for OAT, TAT, and TNT cases.
(d) Plot of $\xi^2$ (dB) as a function of $\chi t$ for OAT, TAT, and TNT cases.
(e) Log-log plot of $\lVert \mathcal{D} \rVert_{\rm F}$ as a function of $N$ for the optimal TAT case. The increasing value with $N$ is merely a random fluctuation on this scale.
(f) Log-log plot of optimal $|\Delta\bm\phi|^2$ as a function of $N$ for OAT, TAT, and TNT cases, compared with the multi-GHZ case $|\Psi\ra$, SQL, and HL. $N$ is fixed at 100 for (b-d). 
}
\label{fig2}
\end{figure*} 

Without loss of generality,
let us analytically examine this result in the limit of $\bm\phi\to0$.
See detailed calculation in App.~\ref{appD}. 
Note that the ensemble of spins exhibits permutation symmetry, represented by the Dicke basis, and can be implemented using collective (global) operators.
At $\bm\phi \to 0$, we have $\bm A \approx \bm J$.
Then, each element of the matrix $\mathcal{D}$ is given by
\begin{align}\label{eq:de}
\mathcal{D}_{\mu\nu} ={\rm Im}[\la\Psi|J_\mu J_\nu |\Psi\ra].
\end{align}
For $\mu = \nu$, we have $\mathcal{D}_{\mu\nu} = 0$.
For $\mu \ne \nu$, we expand Eq.~\eqref{eq:de} into the Dicke basis
$\{|m\ra\}$ as 
\begin{align}\notag
\mathcal{D}_{\mu\nu} =\sum_{m,m',m'' = -J}^J{\rm Im}
[\la\Psi|m\ra\la m|J_\mu|m'\ra\la m' |J_\nu |m''\ra\la m''|\Psi\ra],
\end{align}
where $J = N/2$.
For odd $N$, the nonzero terms are $\mathcal{D}_{yz}$
and $\mathcal{D}_{zy} = - \mathcal{D}_{yz}$, 
which gives 
\begin{align}\label{eq:Dyz}
\notag
\mathcal{D}_{yz} = \frac{J}{2} 
\Big[\la\Psi|-J+1\ra&\la -J |\Psi\ra
-\la\Psi|J-1\ra\la J |\Psi\ra
\Big]\\
&\times \sqrt{J(J+1)-J(J-1)}.
\end{align}
This term is nonzero
when $|\Psi\ra$ is asymmetric under $|\pm J\ra $ and $|\pm (J-1)\ra$, leading to $\lVert \mathcal{D} \rVert_{\rm F} \ne 0$. If $|\Psi\ra$ is symmetric, 
or the asymmetry of $|\Psi\ra$ becomes negligible, e.g., large $N$, then $\mathcal{D}_{yz} = 0$, which results in $\lVert \mathcal{D} \rVert_{\rm F} = 0$.
The analysis is the same for even $N$, yielding the same result for $\mathcal{D}$.

In Fig.~\ref{fig1}b, we plot the probability $P(m) = |\la m|\Psi\ra|^2$ for $\frac{-N}{2} \le m \le \frac{N}{2}$, which represents the projection of the probe state onto the $z$ axis. In this example, the asymmetry is present for $N = 15$ as shown by the purple circles, while it is symmetric for $N = 16$. 
Additional examples with $N = 17, 18$ can be found in App.~\ref{appD}.
Correspondingly, Fig.~\ref{fig1}a indicates that $\lVert \mathcal{D} \rVert_{\rm F} > 0$ at $N = 15$ and $\lVert \mathcal{D} \rVert_{\rm F} \approx 0$ at $N = 16$. 
As $N$ becomes large, the asymmetry becomes negligible, as seen with $N$ = 40 and $N$ = 41 in this example, resulting in $\lVert \mathcal{D} \rVert_{\rm F} \approx 0$.

To understand the symmetry-asymmetry behavior, we next analyze the spin fluctuation, quantified by the variance $\Delta^2$. It is given by $\Delta^2 = \iint Q(\theta, \varphi)(\theta-\bar\theta)^2 (\varphi-\bar\varphi)^2\ d\theta d\varphi$, where $Q(\theta, \varphi) = |\la\Psi|\theta, \varphi\ra|^2$ is the Husimi distribution function. Here, $\bar\theta$ and $\bar\varphi$ denote the mean values of the angles $\theta$ and $\varphi$ in the spherical coordinate system. 
The fluctuation decreases as the number of spins $N$ increases as indicated in Fig.~\ref{fig1}c.

In Fig.~\ref{fig1}c, we compare spin fluctuations for small and large $N$. Apparently,  for small $N$, the large fluctuations in all $|\psi_\mu\ra$ components lead to the overlap and interference, causing deformation and easier symmetry breaking in $|\Psi\ra$.
Conversely, for larger $N$, 
the asymmetry is small and 
becomes negligible.
%
The overlap and interference are visible in the Husimi distribution shown at the bottom of Fig.~\ref{fig1}c, highlighted by the red arrows and circles. 
Notably, the symmetry-asymmetry behavior is not strictly tied to an odd or even number of spins.


Finally, given that the condition is met, we analyze the total variance $|\Delta\bm\phi|^2 = \text{Tr}[\mathcal{C}(\bm\phi)]$, which is now expressed as $\text{Tr}[\mathcal{I}^{-1}]/M$. Figure~\ref{fig1}d illustrates the total variance as a function of $N$, showing a Heisenberg scaling similar to that in Ref. \cite{PhysRevLett.116.030801}.

{\it Quantum-enhanced with a single GHZ probe state}.--- 
In a realistic scenario, assume that 
we can only prepare one component, 
i.e., $|\Psi\ra = |\psi_z\ra = (|\!+\!z\ra + |\!-\!z\ra)/\sqrt{2}$, 
using entanglement amplification techniques \cite{Zhao2021}.
Here, we use $|\pm z\ra$ for brevity instead of $|\lambda_z^{\max (\min)}\ra$.
In this case, the precision increases for estimating $\phi_z$ 
but decreases for $\phi_x$ and $\phi_y$. 
To enhance the precision in all directions, we aim to transform a single GHZ state into a multi-GHZ state using nonlinear squeezing techniques, including OAT \cite{PhysRevA.94.010102,PhysRevA.98.012129,PhysRevA.98.030303,PhysRevA.108.062611}, TAT \cite{PhysRevA.98.030303,PhysRevA.97.043813}, and TNT \cite{PhysRevA.98.030303,PhysRevA.97.053618,PhysRevA.97.043813} transformations. Remarkably, the TAT yields a multi-GHZ-like state, providing high precisions for all phases, similar to the multi-GHZ state.

Particularly, we use squeezing techniques to compress 
the probe state $|\psi_z\ra$, 
following by encoding the phases and reverting (echo), 
as shown in Fig.~\ref{fig2}a.
The nonlinear squeezing methods are represented by the operators
\begin{align}\label{eq:oat}
\notag U_{\rm OAT} &= e^{-it\chi J_x^2}; \;
U_{\rm TAT} = e^{-it\chi (J_x^2 - J_y^2)}; \; \\
U_{\rm TNT} &= e^{-it (\chi J_x^2 - \Omega J_y)} = 
e^{-it\chi(J_x^2 - \frac{N}{\Lambda} J_y)} ,
\end{align}
where $\chi$ represents the magnitude of the spin-spin interaction, $\Omega$ stands for the rate of rotation about the $y$ axis, and we also introduce $\Lambda = N\chi/\Omega$. When $\chi \gg \Omega$ or $\Lambda \gg N$, then TNT simplifies to OAT. Hereafter, we set $\Lambda/N = 0.02$ to investigate the effect of TNT. These transformations are experimentally confirmed \cite{PhysRevA.92.023603,Sorensen2001,Riedel2010,PhysRevLett.87.170402,Gross_2012}.

The estimation scheme is as follows. We first  induce squeezing on the probe state $|\psi_z\ra$ using $U_{\rm k}$ for k = \{OAT, TAT, TNT\}, followed by the phasing unitary $U(\bm\phi) = e^{-iH(\bm\phi)}$. 
We also set $\phi_x =
\phi_y = \phi_z = \phi$ for numerical calculation.
Then, we apply an inverted dynamic $U_{\rm k}^{-r}$, where $r \in \mathbb{R}$ is an arbitrary constant \cite{PhysRevA.97.043813}, resulting in the final state
$
|\psi_z({\bm\phi})\ra = U_{\rm k}^{-r} U({\bm\phi})U_{\rm k}|\psi_z\ra.
$
In general, $U_{\rm k}^{-r}$ does not affect the QFIM as following
\begin{align}\label{eq:QFIM:conTNT}
\notag \mathcal{I} &= 
4
{\rm Re}\bigl[\la\psi_z|U_{\rm k}^\dagger \bm A^\intercal\bm A U_{\rm k}|\psi_z\ra\\
&\hspace{1cm} -\la\psi_z|U_{\rm k}^\dagger \bm A^\intercal U_{\rm k}|\psi_z\ra
\la\psi_z| U_{\rm k}^\dagger \bm A U_{\rm k}|\psi_z\ra\bigr].
\end{align} 
See the proof in App.~\ref{appE}. 


In Fig.~\ref{fig2}c, we plot $|\Delta\bm\phi|^2$ as a function of the squeezing angle $\chi t$ for $N = 100$. We observe that each type of squeezing achieves a minimum value at a certain angle. The OAT result saturates at its optimal angle, the TAT result attains the highest precision at a small optimal angle  $\chi t$, and the TNT result exhibits nonlinear oscillations as $\chi t$ increases.

In Fig.~\ref{fig2}d, we plot the squeezing parameter $\xi^2$ as a function of $\chi t$, following the definition by Kitagawa and Ueda \cite{PhysRevA.47.5138}. 
More details can be found in App.~\ref{appF}.
For single-phase estimations, $\xi^2$ is proportional to the variance, i.e., $\xi^2 \propto |\Delta\phi|^2$ \cite{MA201189,PhysRevA.108.062611}.
In our case, we observe a similar behavior for $\xi^2$ and $|\Delta\bm\phi|^2$, where $\xi^2_{\min}$ aligns with $|\Delta\bm\phi|^2_{\min}$. As $\chi t$ changes continuously, the squeezing parameter tends to increase after reaching its minimum value and does not return to the minimum.

To understand the precision-enhanced mechanism, in Fig.~\ref{fig2}b we examine the Husimi distribution function of the squeezed state at various $\chi t$ points. In the OAT case, the probe state evolves from coherent to squeezing in the $x$-$y$ plane at an angle determined by $\chi t$ \cite{PhysRevA.47.5138}. When $\chi t$ increases, these components stretch and rotate toward the $y$-axis (see also App.~\ref{appG}). When they overlap, interference occurs, resulting in bright and dark points in the Husimi distribution function. Consequently, the squeezed state spreads in both $\pm z$ and $\pm y$ directions, reducing the total variance until it stabilizes as no additional information is added.

Conversely, in the TAT case, the squeezing occurs in the $x$-$y$ plane at $\pm 45$\textdegree \ for $|\!\pm z\ra$, respectively (see App.~\ref{appG}).
As $\chi t$ increases, these components extend in those directions. Due to spherical symmetry, we rotate the state 45\textdegree\ along the $z$ axis, causing it to distribute in both the $\pm x$ and $\pm y$ directions. At the optimal point, the state evenly spreads across all $\pm x, \pm y, \pm z$ directions, resembling the multi-GHZ state $|\Psi\ra$. This causes both $|\Delta\bm\phi|^2$ and $\xi^2$ to reach their minimum values, as indicated by the orange arrows.
More visualization about this case can be found in App.~\ref{appG} and the 
animation (animation.pm4).
%

The explanation for the TNT case is similar to that of OAT.
In this case, the squeezed state distributes along the $\pm z, -x$, and $+y$ axes
and also interfere with each other.

In Fig.~\ref{fig2}e, we analyze the TAT case and calculate the Frobenius norm $\lVert \mathcal{D} \rVert_{\rm F}$ at the optimal $\chi t$ for different $N$, confirming the condition $\lVert \mathcal{D} \rVert_{\rm F} = 0$. 
The increasing value with $N$ is merely a random fluctuation on this scale.
Finally, we examine the variance in Fig.~\ref{fig2}f. 
The variance with TAT reverts to or improves upon that of the $|\Psi\rangle$ scenario, while the variance with OAT only achieves the SQL, and the variance with TNT transitions from SQL to HL.

\begin{figure} [t]
\includegraphics[width=8.6cm]{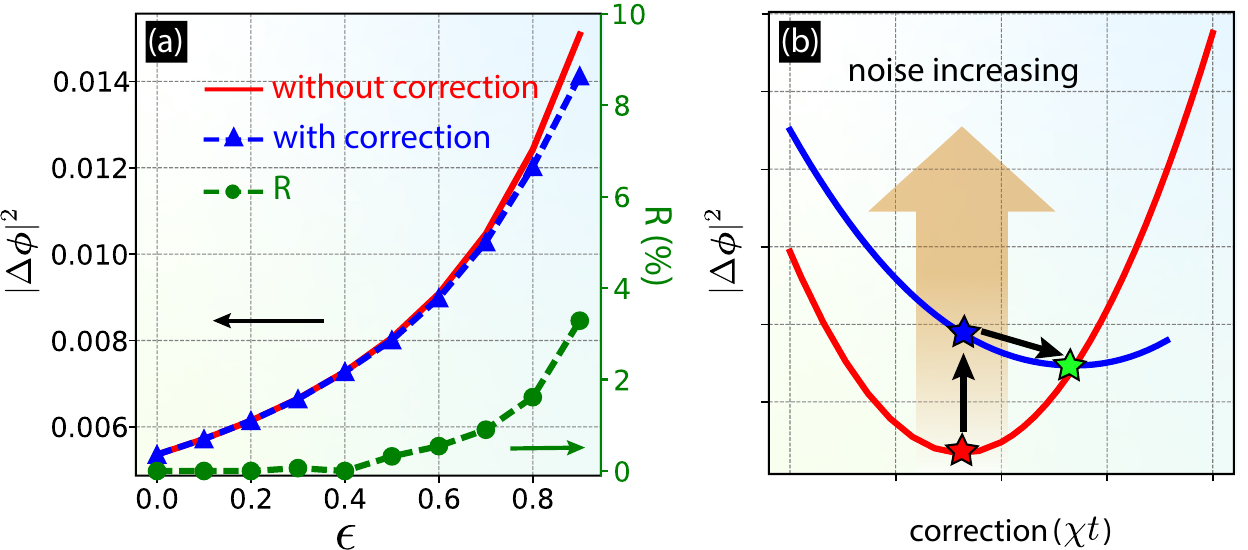}
\caption{(a) Plot of
$|\Delta\bm\phi|^2$ as a function of the noise probability $\epsilon$
for two cases of the correction and without correction. 
The corresponding enhancement ratio $R$ is shown on the right column. For high noise levels, an increase of up to 3\% is observed.
Data represented for $N = 40$.
(b) Illustration demonstrating the squeezing correction.
}
\label{fig3}
\end{figure} 

{\it Quantum-enhanced under noise}.--- 
We next examine the case with dephasing noise during the encoding process. 
Under this noise, a quantum state $\rho$ evolves to
\begin{align}\label{eq:rho_evl}
\rho' = \mathcal{E}_N(\cdots(\mathcal{E}_2(\mathcal{E}_1(\rho))\cdots),
\end{align}
where
$
\mathcal{E}_n(\rho) = (1-\epsilon)\rho + \epsilon a^{(n)}\rho a^{(n)}
$ and 
$a^{(n)} = \epsilon_xJ_x^{(n)} +  \epsilon_yJ_y^{(n)} +  \epsilon_zJ_z^{(n)},
$
satisfies $[a^{(n)}]^2 = \bm I$ \cite{PhysRevA.102.022602}.
Here, $0 \le \epsilon \le 1$ is the noise probability.
See detailed calculations in App.~\ref{appH}.

In Fig.~\ref{fig3}a, we first optimize $\chi t$ at $\epsilon = 0$ to obtain
the minimum $|\Delta\bm\phi|^2$.
After that, we keep $(\chi t)_{\rm opt}$ constance and examine how $|\Delta\bm\phi|^2$ changes with $\epsilon$. The result is shown by the red curve.
As trivial, increasing $\epsilon$ results in an increase in $|\Delta\bm\phi|^2$, which decreases the precision.
Since $(\chi t)_{\rm opt}$ is fixed, we denote this case as ``without correction." 
Next, we attempt to improve the result by optimizing $\chi t$ for each $\epsilon$. Concretely, at each $\epsilon$, we apply the TAT transformation and optimize $\chi t$ to minimize $|\Delta\bm\phi|^2$. 
As depicted in Fig.~\ref{fig3}a with blue triangles, this correction has a minimal impact for small $\epsilon$, but it slightly enhances the precision for larger $\epsilon$.

To quality the enhancement, we define a ratio $R$ as
\begin{align}\label{eq:R}
R = \dfrac{|\Delta\bm\phi|^2_{\rm wo} - |\Delta\bm\phi|^2_{\rm w}}{
|\Delta\bm\phi|^2_{\rm wo} + |\Delta\bm\phi|^2_{\rm w}} \times 100\%,
\end{align}
where ``w/wo" represents ``with/without" correction.
For $\epsilon \leq 0.4$, $R$ remains at zero. 
However, for larger $\epsilon$, $R$ ranges from 0 to slightly over 3\%. 
Although squeezing can enhance precision in noisy conditions, the effect is minimal.

Figure~\ref{fig3}b illustrates the enhancement mechanism. Initially, the red curve depicts the case at $\epsilon = 0$, where we adjust $\chi t$ to find the optimal $(\chi t)_{\rm opt}$, resulting in the best $|\Delta\bm\phi|^2$, indicated by the red star.
In the presence of noise, the red curve shifts upward, indicated by the 
blue line. Maintaining $(\chi t)_{\rm opt}$ leads to an increase in $|\Delta\bm\phi|^2$ (blue star). However, further adjustment of $\chi t$ yields another optimal point, which reduces $|\Delta\bm\phi|^2$ (magenta star).
Although the correction effect is not significant, it can lead to further improvements in squeezing-based corrections.

{\it Conclusion.}---
We studied the mechanism behind the improving precision in multiphase estimations using nonlinear squeezing techniques. We started with a spin ensemble in a GHZ state along a specific axis. We utilized squeezing techniques such as one-axis twisting and twist-and-turn to transform the quantum state to the other axes, thereby increasing precision along those axes. On the other hand, two-axis twisting extended the ensemble to all other axes, resulting in quantum enhancement across all investigated directions. Understanding this mechanism can aid in the design of more effective quantum sensors based on squeezing techniques.

This work is supported by the 
JSPS KAKENHI Grant Number 23K13025.
All numerical computations in this study were done using the tqix code
\cite{HO2021107902,VIET2023108686}.


\bibliography{refs}

\begin{widetext}
\appendix

\setcounter{equation}{0}
\renewcommand{\theequation}{A.\arabic{equation}}
\section{Multiphase estimation with an $N$-identical ensemble}
\label{appA}

Let us consider a set of $d$ parameters represented in a vector $\bm\phi = (\phi_1, \phi_2, \cdots, \phi_d)^\intercal$ that need to be estimated. These parameters are encoded in a probe of $N$ spin-1/2 particles through the Hamiltonian
\begin{align}\label{app:H}
H(\bm\phi) = \bm\phi \cdot \bm H = \sum_{\mu = 1}^d \phi_\mu H_\mu,
\end{align}
where $\bm H = (H_1, H_2, \cdots, H_d)$ represents a set of local Hamiltonians.

To estimate the parameter $\bm\phi$, we follow these steps:
(i) Prepare a probe state $|\Psi\ra$.
(ii) Encode a phase by applying the unitary transformation $U({\bm \phi})$ to get the state $|\Psi({\bm\phi})\ra = U({\bm \phi}) |\Psi\ra$, where $U({\bm \phi}) = e^{-iH(\bm\phi)}$.
(iii) Use a set of POVM $\{\Pi_k\}$ on the resulting state to measure and obtain an outcome $k$, from which we estimate $\bm{\phi}$.

The classical Fisher information matrix (CFIM) is expressed in terms of measurement probabilities as
\begin{align}\label{app:CFIM}
\mathcal{F} = 
\sum_k 
\dfrac{\partial^2}{\partial\bm\phi\partial\bm\phi^\intercal}
\ln P(k|\bm{\phi}),
\end{align}
where
$P(k|\bm{\phi}) = \la\Psi(\bm\phi)|\Pi_k|\Psi(\bm\phi)\ra$
represents the probability of obtaining the outcome $k$.
The quantum Fisher information matrix (QFIM) for a pure state is given by 
\begin{align}\label{app:QFIM}
\mathcal{I}_{\mu\nu} = 
\dfrac{1}{2}
\la\Psi(\bm\phi)|\bigl(L_\mu L_\nu
+L_\nu L_\mu\bigr)|\Psi(\bm\phi)\ra,
\end{align}
where the operator $L$ is defined in the symmetric logarithmic derivative (SLD) as
\begin{align}\label{app:L}
L_{\mu} = 2\bigl(|\partial_{\phi_\mu}\Psi({\bm\phi})\ra\la\Psi({\bm\phi})|
+|\Psi({\bm\phi})\ra\la\partial_{\phi_\mu}\Psi({\bm\phi})|\bigr).
\end{align}
For concreteness, we derive 
\begin{align}\label{eq:der_psi_phi_temp}
\notag |\partial_{\phi_\mu}\Psi({\bm\phi})\ra 
&= \partial_{\phi_\mu}U(\bm\phi)|\Psi\ra\\
\notag &= \partial_{\phi_\mu}
e^{-iH(\bm\phi)}|\Psi\ra\\
\notag &\overset{[33]}{=} -i\int_0^1 due^{-i(1-u)H(\bm\phi)}\textcolor{blue}{[\partial_{\phi_\mu}H(\bm\phi)]}e^{-iuH(\bm\phi)}|\Psi\ra
\\
\notag &= -ie^{-iH(\bm\phi)}\int_0^1 due^{iuH(\bm\phi)}\textcolor{blue}{H_\mu} e^{-iuH(\bm\phi)}|\Psi\ra
\\
&=-iU(\bm\phi)A_\mu|\Psi\ra,
\end{align}
where
\begin{align}\label{app:Aalpha}
A_\mu = \int_0^1du\ e^{iu H(\bm\phi)}H_\mu
e^{-iu H(\bm\phi)},
\end{align}
is a Hermitian operator \cite{10.1063/1.1705306,PhysRevA.90.022117,PhysRevLett.116.030801}.
Then, the SLD \eqref{app:L} and QFIM \eqref{app:QFIM} are explicitly given as
\begin{align}
L_{\mu}&=
2iU(\bm{\phi})\bigl[|\Psi\ra\la\Psi|,
A_{\mu}\bigr]U^\dagger(\bm\phi), \label{app:SLD:con}\\
\mathcal{I}_{\mu\nu} &= 
4{\rm Re}\bigl[\la\Psi|A_\mu A_\nu|\Psi\ra
-\la\Psi|A_\mu|\Psi\ra
\la\Psi|A_\nu|\Psi\ra\bigr].\label{app:QFIM:con}
\end{align}

If we define $\bm A = (A_1, A_2, \cdots, A_d)$ as
$\bm A = \int_0^1du\ e^{iu H(\bm\phi)}\bm H
e^{-iu H(\bm\phi)}$,
then the QFIM \eqref{app:QFIM:con} is recast as
\begin{align}\label{app:QFIM:conn}
\mathcal{I} = 
4{\rm Re}\bigl[\la\Psi|\bm A^\intercal\bm A|\Psi\ra
-\la\Psi|\bm A^\intercal|\Psi\ra
\la\Psi|\bm A|\Psi\ra\bigr].
\end{align}

\setcounter{equation}{0}
\renewcommand{\theequation}{B.\arabic{equation}}
\section{Quantum Cram\'er-Rao bound}\label{appB}

The precision of estimating $\bm\phi$ 
is evaluated by its covariance matrix, 
\begin{align}\label{ap:covar}
\mathcal {C}(\bm\phi) = 
\la {\bm \phi}{\bm \phi}^\intercal \ra-
\la  {\bm \phi} \ra \la  {\bm \phi}^\intercal\ra.
\end{align}
The diagonal elements $[\mathcal{C}(\bm{\phi})]_{\mu\mu}$ represent the variance $\Delta^2\phi_\mu$, while the off-diagonal elements indicate the correlations between different parameters.
The Cram\'er-Rao bounds serve as lower bounds for the covariance matrix, determined by the CFIM and QFIM, as follows
\begin{align}\label{eq:lb}
M\mathcal{C}(\bm{\phi}) \ge {\mathcal F}^{-1}
\ge {\mathcal I}^{-1},
\end{align}
where $M$ represents the repetition of the entire process.
The first inequality represents the classical Cram\'er-Rao bound (CCRB), while the second one is known as the quantum Cram\'er-Rao bound (QCRB).
Since we are solely focusing on quantum-enhanced measurement, we choose $M = 1$.
The total variance of all phases is then given by
\begin{align}\label{app:covar}
|\Delta\bm\phi|^2 = \sum_{\mu = 1}^d \Delta^2\phi_\mu
={\rm Tr}[\mathcal{C}(\bm\phi)].
\end{align}
When the QCRB is satisfied, 
it implies that
$|\Delta\bm\phi|^2 = {\rm Tr}[\mathcal{I}^{-1}]$.

\setcounter{equation}{0}
\renewcommand{\theequation}{C.\arabic{equation}}
\section{QCRB saturating} \label{appC}

We explore the necessary condition for the saturation of the QCRB, which is determined by the expectation value of the commutator of the SLD \cite{Matsumoto_2002}
\begin{align}\label{app:condi}
\la\Psi({\bm\phi})|L_\mu L_\nu - L_\nu L_\mu|\Psi({\bm\phi})\ra
=0.
\end{align}
We begin by deriving $L_\mu L_\nu$ as
\begin{align}\label{app:LmuLnu}
\notag L_\mu L_\nu &= -4U(\bm{\phi})
	\textcolor{blue}{\bigl[|\Psi\ra\la\Psi|,A_{\mu}\bigr]} \cdot
	\textcolor{magenta}{\bigl[|\Psi\ra\la\Psi|,A_{\nu}\bigr]}
	U^\dagger(\bm\phi)\\
\notag &= -4U(\bm{\phi})
	\textcolor{blue}{\bigl(|\Psi\ra\la\Psi|A_{\mu} - A_{\mu}|\Psi\ra\la\Psi|\bigr)} \cdot
	\textcolor{magenta}{\bigl(|\Psi\ra\la\Psi|A_{\nu} - A_{\nu}|\Psi\ra\la\Psi|\bigr)}
	U^\dagger(\bm\phi) \\
&= -4U(\bm{\phi})
	\bigl(
	\textcolor{blue}{|\Psi\ra\la\Psi|A_{\mu}}
	\textcolor{magenta}{|\Psi\ra\la\Psi|A_{\nu}}
		- 
	\textcolor{blue}{|\Psi\ra\la\Psi|A_{\mu}}
	\textcolor{magenta}{A_{\nu}|\Psi\ra\la\Psi|}
		- 
	\textcolor{blue}{A_{\mu}|\Psi\ra}
	\textcolor{magenta}{\la\Psi|A_{\nu}}
		+ 
	\textcolor{blue}{A_{\mu}|\Psi\ra\la\Psi|}
	\textcolor{magenta}{A_{\nu}|\Psi\ra\la\Psi| }\bigr)
	U^\dagger(\bm\phi).
\end{align}
Then, we have
\begin{align}\label{app:Exp_LmuLnu}
\la\Psi(\bm\phi)| L_\mu L_\nu |\Psi(\bm\phi)\ra = 
	4\big(
	\la\Psi|A_{\mu}A_{\nu}|\Psi\ra-
	\la\Psi|A_{\mu}|\Psi\ra\la\Psi|A_{\nu}|\Psi\ra
	\big).
\end{align}
The term $\la\Psi(\bm\phi)| L_\nu L_\mu |\Psi(\bm\phi)\ra$
is obtained by interchanging $\mu \leftrightarrow \nu$,
resulting in $\la\Psi(\bm\phi)| L_\nu L_\mu |\Psi(\bm\phi)\ra
= [\la\Psi(\bm\phi)| L_\mu L_\nu |\Psi(\bm\phi)\ra]^*$.
The right hand side of \eqref{app:condi} yields 
\begin{align}\label{app:RHS}
\la\Psi({\bm\phi})|L_\mu L_\nu - L_\nu L_\mu|\Psi({\bm\phi})\ra
= 8 {\rm Im} [\la\Psi|A_\mu A_\nu|\Psi\ra].
\end{align}
Finally, we recast the condition \eqref{app:condi} as
\begin{align}\label{app:condition}
{\rm Im}[\la\Psi|A_\mu A_\nu|\Psi\ra] = 0.
\end{align}
For all $\mu, \nu \in (1, 2, \cdots, d)$, we recall
Eq.~\eqref{app:condition} in a matrix form as
\begin{align}\label{app:conditionMat}
\mathcal{D} \equiv {\rm Im}[\la\Psi|\bm A^\intercal\bm A|\Psi\ra] = 0,
\end{align}
where $\bm A = (A_1, A_2, \cdots, A_d)$.

In our model, we first note that for individual GHZ states,
i.e., $|\Psi\ra = |\psi_\mu \ra, \mu = x, y, z$,
condition~\eqref{app:condition} is always satisfied 
regardless of the number of spins. 
This result is due to the symmetry presented in the bases
$|\lambda_\mu^{\rm max}\ra$ and $|\lambda_\mu^{\rm min}\ra$.
When $|\Psi\ra$ is a summation of all components, as shown in Eq.~\eqref{eq:psi}, and with a large value of $N$, the asymmetry in $|\Psi\ra$ is negligible. 
Thus, the condition~\eqref{app:conditionMat} still holds, 
as demonstrated in the main text.

When the condition \eqref{app:conditionMat} is met, it also requires a set of POVM that enables the achievement of the QCRB (the sufficient condition). One corresponding set of optimal POVM
is given by $\Pi_{k=1} = |\Psi({\bm\phi})\ra\la\Psi({\bm\phi})|$ and $\Pi_{k\ne 1} = |\Phi_k\ra \la\Phi_k|$, where $\Pi_{k=1} +\sum_{k\ne 1} \Pi_{k\ne 1} = I$. We emphasize that this set of POVM is determined at a fixed point of $\bm \phi$. However, with such optimal POVMs, the QCRB can be achieved \cite{PhysRevLett.111.070403}.


\setcounter{equation}{0}
\renewcommand{\theequation}{D.\arabic{equation}}
\section{QCRB saturating under the multi-GHZ case} \label{appD}

Now, we analytically demonstrate the saturating of condition \eqref{app:conditionMat} in the limit of $\bm\phi\to 0$. 
We emphasize that for
a small phase $\bm\phi$, 
numerical results indicate that $\mathcal{D} \approx 0$. 
However, we do not prove this case here.)
Starting from $\bm A$, we have
\begin{align}\label{eq:Aappd}
\notag \bm A &= \int_0^1du\ e^{iu H(\bm\phi)}\bm Je^{-iu H(\bm\phi)}\\
\notag &\approx \int_0^1du\ \big(\bm I + iuH(\bm\phi) + O(\bm\phi^2) \big)\bm J
\big(\bm I - iuH(\bm\phi) + O(\bm\phi^2) \big)\\
&= \bm J + iu \big[H(\bm\phi),\bm J\big] + O(\bm\phi^2) .
\end{align}
For $\bm\phi = 0$, we have $\bm A \approx \bm J$.
Then, each element of matrix $\mathcal{D}$ is given by
\begin{align}\label{appeq:de}
\mathcal{D}_{\mu\nu} ={\rm Im}[\la\Psi|J_\mu J_\nu |\Psi\ra].
\end{align}

We identify all nonzero terms $\mathcal{D}_{\mu\nu} \forall \mu, \nu$, and examine the conditions under which they become zero.
Utilizing the permutation symmetry of the probe system \cite{PhysRevA.102.022602}, and in the absence of noise, the probe can be represented in the Dicke basis ${|J,m\ra}$, where $J = N/2$, and $-J \le m \le J$, or denoted as ${|m\ra}$ for short. The angular momentum operators are defined by spin-$J$ operators, where
\begin{align}\label{app:spino}
\la m'|J_x|m\ra &= \dfrac{1}{2}(\delta_{m',m+1}+\delta_{m'+1,m})\sqrt{J(J+1)-mm'},\\
\la m'|J_y|m\ra &= \dfrac{-i}{2}(\delta_{m',m+1}-\delta_{m'+1,m})\sqrt{J(J+1)-mm'},\\
\la m'|J_z|m\ra &= \delta_{m',m}m.
\end{align}

We expand Eq.~\eqref{appeq:de} using the Dicke basis as
\begin{align}\notag
\mathcal{D}_{\mu\nu} ={\rm Im}\sum_{m,m',m''}
\la\Psi|m'\ra\la m'|J_\mu|m\ra\la m |J_\nu |m''\ra\la m''|\Psi\ra.
\end{align}
For $\mu = \nu$, it is evident that $\mathcal{D}_{\mu\nu} = 0$. When $\mu \neq \nu$, we distinguish between two cases: odd $N$ and even $N$. In the case of odd $N$, the nonzero terms are
$\mathcal{D}_{yz}$ and $\mathcal{D}_{zy} = - \mathcal{D}_{yz}$, where
\begin{align}\notag
\mathcal{D}_{yz} &= 
{\rm Im}\sum_{m,m',m''}
[\la\Psi|m'\ra\la m'|J_y|m\ra\la m |J_z |m''\ra\la m''|\Psi\ra],\\
\notag &= {\rm Im}\sum_{m,m'}
m\la\Psi|m'\ra\la m'|J_y|m\ra \la m|\Psi\ra, \text {(where use used $m'' = m$)}\\
&=\sum_m\frac{m}{2}\Big[\textcolor{blue}{\la\Psi|m+1\ra\la m |\Psi\ra \sqrt{J(J+1)-m(m+1)}}
- \textcolor{magenta}{\la\Psi|m-1\ra\la m |\Psi\ra\sqrt{J(J+1)-m(m-1)}}\Big].
\end{align}
The blue and magenta terms will cancel out for $m = \pm k, \forall k\in \{1/2, \cdots, J-1\}$. The only nonzero terms occur when $m = \pm J$, which is
\begin{align}\label{app:nonz}
\notag\mathcal{D}_{yz} &= 
\frac{-J}{2}\textcolor{magenta}{\la\Psi|J-1\ra\la J |\Psi\ra \sqrt{J(J+1)-J(J-1)}}
+\frac{J}{2} \textcolor{blue}{\la\Psi|-J+1\ra\la -J |\Psi\ra\sqrt{J(J+1)+J(-J+1)}}\\
&= \frac{J}{2} 
\Big[\textcolor{blue}{\la\Psi|-J+1\ra\la -J |\Psi\ra}
-\textcolor{magenta}{\la\Psi|J-1\ra\la J |\Psi\ra}
\Big]
\sqrt{J(J+1)-J(J-1)}.
\end{align}

For small $N$, it is evident that $\mathcal{D}_{yz} \neq 0$ when the probe state $|\Psi\ra$ is asymmetry under $|\pm J\ra $ and $|\pm (J-1)\ra$. Inversely, if $|\Psi\ra$ is symmetric, the term $[\star]$ in \eqref{app:nonz} vanishes, leading to $\mathcal{D}_{yz} = 0$. As $N$ grows large, the asymmetry of $|\Psi\ra$ becomes negligible, resulting in $\mathcal{D}_{yz} = 0$.


In the case of even $N$, the nonzero terms are $\mathcal{D}_{xz}$ and $\mathcal{D}_{zx} = - \mathcal{D}_{xz}$. These can be calculated using the same method and exhibit the same behavior.

In Fig.~\ref{fig4}, we plot the probability $P(m) = |\langle m|\Psi \rangle|^2$ for \(N\) ranging from 15 to 18 and their corresponding \(\lVert \mathcal{D} \rVert_{\rm F}\). Within this range, only the wavefunction with \(N = 16\) is symmetric, resulting in \(\lVert \mathcal{D} \rVert_{\rm F} \propto 10^{-15}\). The other cases exhibit asymmetry, leading to nonzero \(\lVert \mathcal{D} \rVert_{\rm F}\).
\begin{figure*} [t!]
\includegraphics[width=12.6cm]{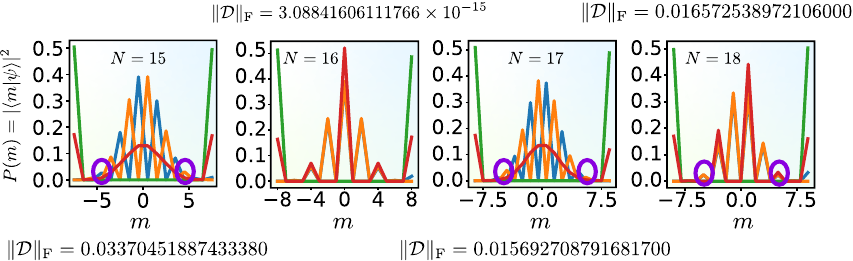}
\caption{Plot of $P(m) = |\langle m|\Psi\rangle|^2$ for \(N\) from 15 to 18 and their corresponding \(\lVert \mathcal{D} \rVert_{\rm F}\) talking from Fig.~\ref{fig1}a for \(\phi = 0.01\).
}
\label{fig4}
\end{figure*}

\setcounter{equation}{0}
\renewcommand{\theequation}{E.\arabic{equation}}
\section{Quantum Fisher Information matrix under 
squeezing}\label{appE}

To enhance the precision of $|\Delta\bm\phi|^2$, we employ different squeezing methods, including one-axis twisting (OAT), two-axis twisting (TAT), and twist-and-turn (TNT) transformations. Let $U_{\rm k}$ represent the unitary transformation associated with each method, where k = OAT, TAT, or TNT.

We start by subjecting the probe state \(|\psi_z\ra\) to squeezing using \(U_{\rm k}\). Next, we apply the phasing unitary \(U(\bm\phi) = e^{-iH(\bm\phi)}\) and an inverted dynamic \(U_{\rm k}^{-r}\), where \(r\) is any real constant \cite{PhysRevA.97.043813}. This process results in the final state
\begin{align}\label{app:psiphi}
|\psi_z({\bm\phi})\ra = U_{\rm k}^{-r} U({\bm\phi})U_{\rm k}|\psi_z\ra.
\end{align}
It is important to note that \(U_{\rm k}^{-r}\) can also be represented by an identity matrix \(\bm{I}\) or \(U_{\rm k}^\dagger\) \cite{PhysRevA.97.053618}.

We first calculate the derivative
\begin{align}\label{eq:der_psi_phi_temp_appA}
\notag |\partial_{\phi_\mu}\psi_z({\bm\phi})\ra 
&= U_{\rm k}^{-r} 
\textcolor{blue}{\bigl[\partial_{\phi_\mu}U(\bm\phi)\bigr]}U_{\rm k}|\psi_z\ra\\
\notag &= U_{\rm k}^{-r}
\textcolor{blue}{\bigl[\partial_{\phi_\mu}e^{-iH(\bm\phi)}\bigr]}
U_{\rm k}|\psi_z\ra\\
\notag &= -iU_{\rm k}^{-r}
\textcolor{blue}{\Bigl[\int_0^1 due^{-i(1-u)H(\bm\phi)}[\partial_{\phi_\mu}H(\bm\phi)]e^{-iuH(\bm\phi)}\Bigr]}
U_{\rm k}|\psi_z\ra
\\
\notag &= -iU_{\rm k}^{-r}
\textcolor{blue}{e^{-iH(\bm\phi)}\Bigl[\int_0^1 due^{iuH(\bm\phi)}H_\mu e^{-iuH(\bm\phi)}\Bigr]}
U_{\rm k}|\psi_z\ra
\\
&=-iU_{\rm k}^{-r}\textcolor{blue}{U(\bm\phi)A_\mu} U_{\rm k}|\psi_z\ra.
\end{align}
We next calculate the SLD \eqref{app:L}:
\begin{align}\label{eq:L_appA}
\notag L_{\mu} &= 2\Bigl(
\textcolor{blue}{|\partial_{\phi_\mu}\psi_z({\bm\phi})\ra}
\la\psi_z({\bm\phi})|
+|\psi_z({\bm\phi})\ra
\textcolor{magenta}{\la\partial_{\phi_\mu}\psi_z({\bm\phi})|}\Bigr)\\
\notag&=2
\Bigl(
\textcolor{blue}{-iU_{\rm k}^{-r}U(\bm\phi)A_\mu U_{\rm k}|\psi_z\ra}
\la\psi_z|U_{\rm k}^\dagger U^\dagger(\bm\phi)[U_{\rm k}^{-r}]^\dagger
+iU_{\rm k}^{-r} U({\bm\phi})U_{\rm k}|\psi_z\ra
\textcolor{magenta}{\la\psi_z|U_{\rm k}^\dagger A_\mu U^\dagger(\bm\phi)[U_{\rm k}^{-r}]^\dagger}
\Bigr)
\\
&=2iU_{\rm k}^{-r}U(\bm\phi)
\Bigl[U_{\rm k}|\psi_z\ra\la\psi_z|U_{\rm k}^\dagger,A_\mu\Bigr]
U^\dagger(\bm\phi)[U_{\rm k}^{-r}]^\dagger.
\end{align}
We now calculate $L_\mu L_\nu$ as
\begin{align}\label{eq:LaLb_appA}
L_\mu L_\nu &= 
-4U_{\rm k}^{-r}U(\bm\phi)
\Bigl[U_{\rm k}|\psi_z\ra\la\psi_z|U_{\rm k}^\dagger,A_\mu\Bigr]\cdot
\Bigl[U_{\rm k}|\psi_z\ra\la\psi_z|U_{\rm k}^\dagger,A_\nu\Bigr]
U^\dagger(\bm\phi)[U_{\rm k}^{-r}]^\dagger.
\end{align}
We next evaluate the term 
$\la\psi_z({\bm\phi})|L_\mu L_\nu|\psi_z({\bm\phi})\ra$ as
\begin{align}\label{app:spi_LaLb_appA_psi_appA}
\la\psi_z({\bm\phi})|L_\mu L_\nu|\psi_z({\bm\phi})\ra
= 4
\Bigl(
\la\psi_z|U_{\rm k}^\dagger A_\mu A_\nu U_{\rm k}|\psi_z\ra
-
\la\psi_z|U_{\rm k}^\dagger A_\mu U_{\rm k}|\psi_z\ra
\la\psi_z|U_{\rm k}^\dagger A_\nu U_{\rm k}|\psi_z\ra
\Bigr),
\end{align}
and 
$
\la\psi_z({\bm\phi})|L_\nu L_\mu|\psi_z({\bm\phi})\ra
= [\la\psi_z({\bm\phi})|L_\mu L_\nu|\psi_z({\bm\phi})\ra]^*
$.
Finally, we calculate the element of QFIM \eqref{app:QFIM} as
\begin{align}\label{app:QFIM_appA}
\notag \mathcal{I}_{\mu\nu} &= 
\dfrac{1}{2}
\la\psi_z({\bm\phi})|\bigl(L_\mu L_\nu
+L_\nu L_\mu\bigr)|\psi_z({\bm\phi})\ra\\
&=4{\rm Re}\bigl[\la\psi_z|U_{\rm k}^\dagger A_\mu A_\nu U_{\rm k}|\psi_z\ra
-\la\psi_z|U_{\rm k}^\dagger A_\mu U_{\rm k}|\psi_z\ra
\la\psi_z| U_{\rm k}^\dagger A_\nu U_{\rm k}|\psi_z\ra\bigr].
\end{align}
%
And the full QFIM yields 
\begin{align}\label{app:QFIM_appA_F}
\mathcal{I} = 
4{\rm Re}\bigl[\la\psi_z|U_{\rm k}^\dagger \bm A^\intercal \bm A U_{\rm k}|\psi_z\ra
-\la\psi_z|U_{\rm k}^\dagger \bm A^\intercal U_{\rm k}|\psi_z\ra
\la\psi_z| U_{\rm k}^\dagger \bm A U_{\rm k}|\psi_z\ra\bigr].
\end{align}

\setcounter{equation}{0}
\renewcommand{\theequation}{F.\arabic{equation}}
\section{Spin squeezing parameter}\label{appF}

The definition of spin squeezing varies depending on the context of different authors, leading to multiple definitions in the literature \cite{PhysRevA.47.5138,PhysRevA.50.67,Sorensen2001,MA201189}. 
One early definition of the spin squeezing parameter is derived from the uncertainty relation for the angular momentum operators
\begin{align}\label{app:sq1}
(\Delta J_x)^2(\Delta J_y)^2 
\ge \dfrac{|\la J_z \ra|^2}{4},
\end{align}
which is known as the Heisenberg uncertainty relation,
i.e., $\xi_{\rm H}^2 = N\frac{(\Delta \bm J_x)^2}
{|\la\bm J_z\ra|^2},$
where H stands for 
the Heisenberg. 
Two well-known spin squeezing parameters originate from the works of Kitagawa and Ueda \cite{PhysRevA.47.5138} and Wineland \cite{PhysRevA.50.67}. According to Kitagawa and Ueda, the spin-squeezing state (SSS) redistributes quantum fluctuations between two noncommuting observables while maintaining the minimum uncertainty product \cite{PhysRevA.78.052101, PhysRevA.81.032104}. Unlike bosonic systems, where the variance is uniform in any direction for a bosonic coherent state, in a coherent spin state, the variance of spin operators depends on $\bm n$, with a predefined direction known as the mean-spin direction (MSD) \cite{PhysRevA.78.052101,PhysRevA.81.032104}.
The corresponding squeezing parameter is given by
\begin{align}\label{eq:xi_S_2}
    \xi^2_{\rm S}=
     \frac{2}{N}
     \Big[\la \bm J^2_{\bm n_2}+\bm J^2_{\bm n_3}\ra
     \pm \sqrt{
     \la \bm J^2_{\bm n_2}-\bm J^2_{\bm n_3}\ra^2
     +4{\rm cov}^2(\bm J_{\bm n_2},\bm J_{\bm n_3})}
     \Big],
\end{align}
where
${\bm n}_2 = (-\sin\phi,\cos\phi,0), 
{\bm n}_3 = (\cos\theta \cos\phi,
 \cos\theta \sin\phi,-\sin\theta)$
and 
\begin{align}
\notag   \theta = \arccos \left( 
   \frac{\la J_z\ra}{|\bm{J}|} 
   \right)
   \text{, }
   \phi =  
        \begin{cases}
            \arccos
            \left( 
            \frac{\la J_x \ra}
            {|\bm{J}\sin\theta|} 
            \right)    
   & \text{if } \la J_y \ra > 0, \\
   2\pi - \arccos\left( 
   \frac{\la J_x \ra}
   {|\bm{J}\sin\theta|} 
   \right) 
   & \text{if } 
   \la J_y \ra \leq 0 ,
   \end{cases}
\end{align}
with 
$
  |\bm{J}| = \sqrt{ \la J_x \ra^2 +
  \la J_y \ra^2 + \la J_z \ra^2},
$
and the covariant cov$(\bm J_{\bm n_2},
\bm J_{\bm n_3})$ is given by
\begin{align}\label{eq:cov}
    {\rm cov}(\bm J_{\bm n_2},\bm J_{\bm n_3}) 
    = \frac{1}{2} 
    \la 
    [\bm J_{\bm n_2},\bm J_{\bm n_3}]_+ 
    \ra - 
    \la \bm J_{\bm n_1} \ra
    \la \bm J_{\bm n_2} \ra. 
\end{align}

The Wineland squeezing parameter 
is defined by
\cite{PhysRevA.50.67}
\begin{align}\label{eq:xi_R_2}
     \xi^2_{\rm R}=
     \left(\ \frac{N}
     {2|\la \bm{J} \ra|}
     \right)^2\xi^2_{\rm S}.
\end{align}
When the squeezing parameter $\xi^2 < 1$, the system state is squeezed. Experiments conducted with cold or room-temperature atomic ensembles \cite{PhysRevLett.83.1319,PhysRevLett.109.253605} have achieved up to -20 dB of spin squeezing \cite{Hosten2016}, and this can be further enhanced by employing cavities \cite{PhysRevLett.104.073602}.

\setcounter{equation}{0}
\renewcommand{\theequation}{G.\arabic{equation}}
\section{On the squeezing mechanism for quantum-enhanced metrology} \label{appG}

\begin{figure*} [t!]
\includegraphics[width=6.6cm]{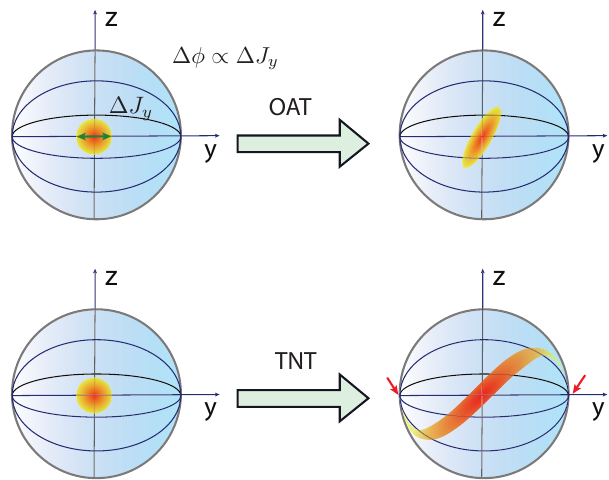}
\caption{
Visualization of a probe state undergoing different squeezing transformations. Initially, the probe state is prepared coherently along the $x$-axis. The standard deviation of the estimated phase $\Delta\phi$ is directly linked to the uncertainty $\Delta S_y$. With the OAT transformation (and similarly with TAT), the uncertainty $\Delta S_y$ reduces, consequently reducing $\Delta\phi$, which enhancing the sensitivity. Similarly, with the TNT transformation, the coherent state extends along the $y$-axis as indicated by the red arrows, resembling a GHZ state, thereby enhancing the sensitivity. 
}
\label{fig5}
\end{figure*}

For single-phase estimation, let us assume we aim to estimate $\phi$ encoded in the unitary operation $U(\phi) = e^{-i\phi J_y}$, representing a rotation around the $y$-axis. To estimate $\phi$, we use a quantum probe initially prepared in the coherent state along the $x$-axis, as illustrated in Fig.~\ref{fig5}.
The sensitivity of the estimation is given via the standard deviation
$\Delta\phi$ as
\begin{align}\label{app:Deltaphi}
\Delta\phi = \dfrac{\Delta J_y}{\partial_\phi\la J_y\ra}\Big|_{\phi = 0}.
\end{align}
In principle, to improve the sensitivity, 
we need to reduces the standard deviation
$\Delta\phi$, such as reduces  $\Delta J_y$.
One approach is employing the one-axis twisting (OAT) transformation along the $z$-axis, denoted by $U_{\rm OAT} = e^{-it\chi J_z^2}$. As depicted in Fig.~\ref{fig5}, this squeezing operation decreases $\Delta J_y$, thereby enhancing sensitivity. Similar effects can be achieved with the twist-and-turn (TAT) transformation, albeit in different directions. The twist-and-turn (TNT) transformation $U_{\rm TNT} = e^{-it(\chi J_z^2-\Omega J_x)}$ offers the most significant enhancement. This transformation not only twists around the $z$ axis but also turns around the $x$ axis. Consequently, the initial state is squeezed into the $|+y\ra$ and $|-y\ra$ components, resembling the GHZ state along the $y$ axis, denoted by $|\psi_y\ra$. As known, $|\psi_y\ra$ allows for attaining the Heisenberg limit for estimating $\phi_y$.

Similarly, we discuss our case of 3D-phase estimation. Since the estimated phases are asymmetric in $x, y, z$, we start with an arbitrary GHZ state, such as $|\psi_z\ra$. We derive $|\psi_z\ra = \frac{1}{\sqrt{2}}(|\!\!+\!\!z\ra + |\!\!-\!\!z\ra$, as shown in the left of Fig.~\ref{fig6} with top-bottom view.

In the context of OAT as depicted in Fig.~\ref{fig6}, the component $|\!\!+z\ra$ undergoes squeezing in quadrants I-III of the $x$-$y$ plane, while the state $|\!\!-z\ra$ undergoes squeezing in quadrants II-IV. As the squeezing angle $\chi t$ increases, these two components stretch and shift toward the $y$ axis, as indicated by the red arrows. Eventually, they overlap and interfere, resulting in both constructive and destructive interference. Consequently, the state distribution extends in both the $\pm z$ and $\pm y$ directions, enhancing the prediction.

When subject to TAT squeezing, the $|+z\ra$ component is squeezed along the 45\textdegree\ quadrants I-III, while its counterpart is squeezed along the 45\textdegree\ quadrants II-IV. In spherical coordinates, a 45\textdegree\ rotation around the $z$-axis can be used to adjust this squeezing into the $\pm x$ and $\pm y$ directions. As a result, the squeezed state is distributed in all directions $\pm x$, $\pm y$, and $\pm z$. This provides the best metrological precision, equivalent to using the $|\Psi\ra$ state.

For further details about this case, refer to the Husimi visualization in Fig.~\ref{fig7}. In this figure, we can see that, the more we squeeze,
the more these components stretch out the $x$ and $y$ axis. 
The maximum precision (minimum total variance) corresponds to the case when 
the distribution to the $x, y, z$ axises are equal.

\begin{figure*} [t!]
\includegraphics[width=14.6cm]{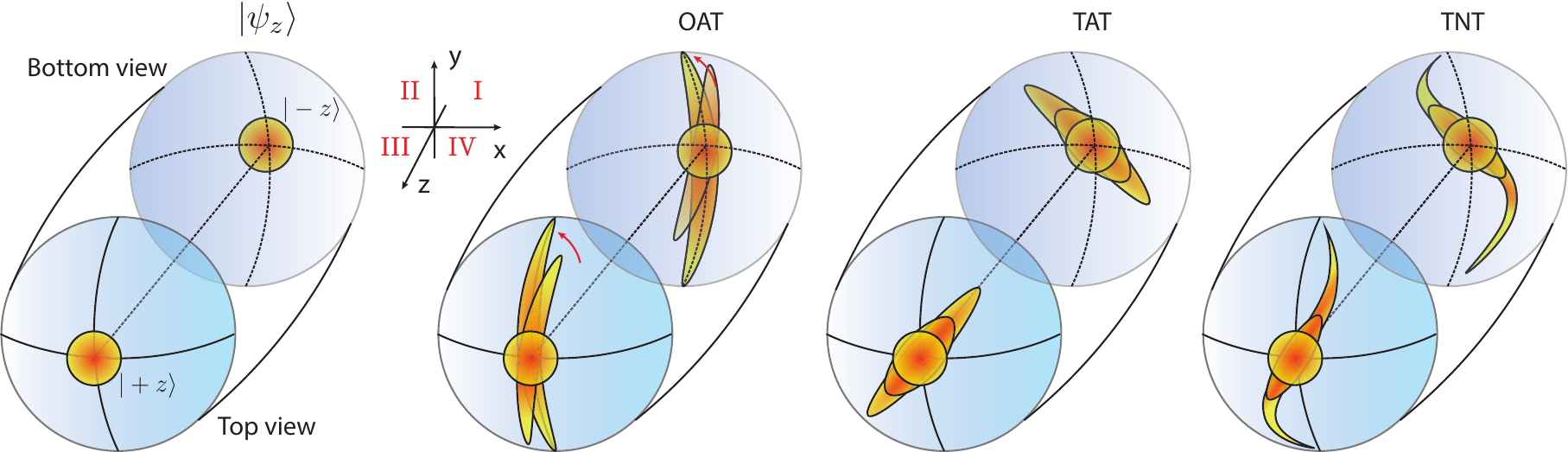}
\caption{
Top-bottom view of the quantum probe state $|\psi_z\ra = (|\!\!+\!\!z\ra + |\!\!-\!\!z\ra)/\sqrt{2}$, 
and its evolved states under various squeezing transformations.
}
\label{fig6}
\end{figure*} 

\begin{figure*} [t!]
\includegraphics[width=15.6cm]{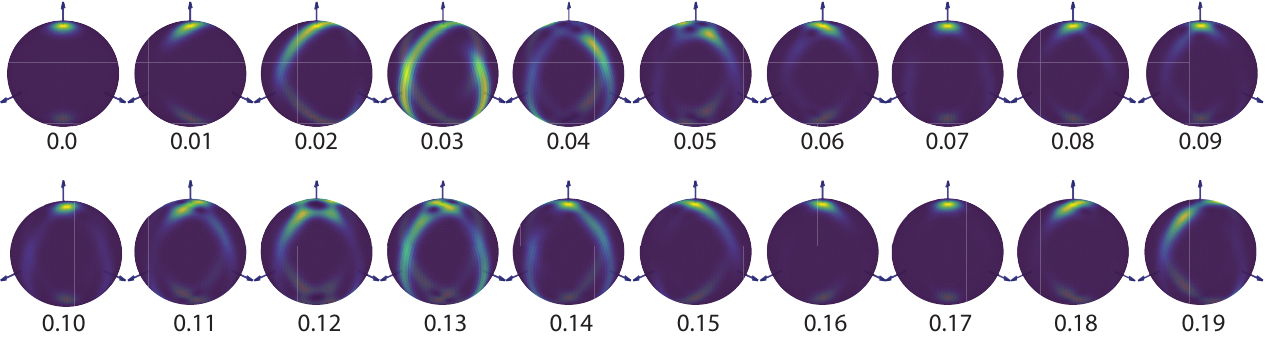}
\caption{
Visualization the Husimi functions 
of $|\psi_z\ra$ under the TAT transformation for
different values of $\chi t$ as shown in the figure.
We fixed $N = 100$.
}
\label{fig7}
\end{figure*} 

Finally, for TNT, we observe squeezing in the $\pm z$ direction and also in the $-x$ and $+y$ directions. This makes it better than OAT but not as effective as TAT.

\setcounter{equation}{0}
\renewcommand{\theequation}{H.\arabic{equation}}
\section{Multiphase estimation under decoherence} \label{appH}
We investigate the scenario where our system is affected by dephasing noise. Initially, we examine the probe state as
\begin{align}\label{eq:rho}
\rho = U_k|\psi_z\ra\la\psi_z|U_k^\dagger.
\end{align}
The $n^{\text{th}}$ particle undergoes a small change
\begin{align}\label{eq:an}
a^{(n)} = \epsilon_xJ_x^{(n)} +  \epsilon_yJ_y^{(n)} +  \epsilon_zJ_z^{(n)},
\end{align}
which satisfies $[a^{(n)}]^2 = \bm I$ \cite{PhysRevA.102.022602}.
Here, we fix $\epsilon_x = \epsilon_y = \epsilon_z = const$.
Under this noise, the quantum state evolves to
\begin{align}\label{eq:rho'}
\rho' = \mathcal{E}_N(\cdots(\mathcal{E}_2(\mathcal{E}_1(\rho))\cdots),
\end{align}
where
\begin{align}\label{eq:epsilon}
\mathcal{E}_n(\rho) = (1-\epsilon)\rho + \epsilon a^{(n)}\rho a^{(n)},
\end{align}
and $0 \le \epsilon \le 1$ is the noise probability.
We now calculate $\rho'$ explicitly: 
\begin{align}\label{eq:epN}
 \mathcal{E}_1(\rho) &= (1-\epsilon)\rho + \epsilon a^{(1)}\rho a^{(1)},\\
\notag \mathcal{E}_2(\textcolor{blue}{\mathcal{E}_1(\rho)}) &= (1-\epsilon)
\textcolor{blue}{[(1-\epsilon)\rho + \epsilon a^{(1)}\rho a^{(1)}]}
+\epsilon a^{(2)}\textcolor{blue}{[(1-\epsilon)\rho + \epsilon a^{(1)}\rho a^{(1)}]}a^{(2)}, \\
&=(1-\epsilon)^2\rho+\epsilon(1-\epsilon)(a^{(1)}\rho a^{(1)}
+ a^{(2)}\rho a^{(2)})
+\epsilon^2a^{(2)}a^{(1)}\rho a^{(1)}a^{(2)},
\end{align}
and so on. Finally, we have
\begin{align}\label{eq:rho'full}
\notag\rho' &= (1-\epsilon)^N\rho+\epsilon(1-\epsilon)^{N-1}\sum_{k=1}^Na^{(k)}\rho a^{(k)} \\
\notag&+\epsilon^2(1-\epsilon)^{N-2}\sum_{k=1}^{N}
\Pi_{l=N, l \ne k}^1 a^{(l)}\rho \Pi_{l=1, l \ne k}^N a^{(l)} \\
\notag&+\cdots \\
&+\epsilon^Na^{(N)}\cdots a^{(2)} a^{(1)}\rho a^{(1)}a^{(2)}\cdots a^{(N)}.
\end{align}
The final probe state is given by
\begin{align}\label{eq:rho_out}
\rho_{\rm out} = U(\bm{\phi})\rho'U^\dagger(\bm {\phi}).
\end{align}
The QFIM is given as
\begin{align}\label{eq:QFIM}
\mathcal{I}_{\mu\nu} = 
{\rm Re}\bigl[{\rm Tr}[\rho_{\rm out} L_\mu L_\nu]\bigr],
\end{align}
where 
$L_\mu = 2\int_0^\infty dt e^{-\rho_{\rm out}t}\bigl(\partial_{\phi_\mu}\rho_{\rm out}\bigr)e^{-\rho_{\rm out}t}$ is the SLD.

\end{widetext}

\end{document}